\documentclass{jfp1} 
\usepackage{colortbl}
\usepackage{enumerate}
\usepackage{natbib}
\usepackage{hyperref}
\usepackage{url}
\usepackage{graphics}
\usepackage{mathpartir}
\usepackage{placeins}
\makeatletter
\expandafter\let\csname equation*\endcsname\@undefined
\expandafter\let\csname endequation*\endcsname\@undefined
\let\proof\@undefined
\let\endproof\@undefined
\expandafter\let\csname proof*\endcsname\@undefined
\expandafter\let\csname endproof*\endcsname\@undefined
\makeatother
\usepackage{amsmath}
\usepackage{amssymb}
\usepackage{fdsymbol}
\usepackage{mathtools}
\usepackage{latexsym}
\usepackage{oktheorem}
\usepackage{tikz}
\newcommand{\Eff}{\emph{Eff}}

\newcommand{\cond}[1]{\quad(#1)}

\newcommand{\bnfis}{\mathrel{\;{:}{:}\!=}\;}
\newcommand{\bnfor}{\mathrel{\;\big|\;}}
\newcommand{\defeq}{\mathrel{\;\stackrel{\text{def}}{=}\;}}

\newcommand{\kord}[1]{\mathbf{#1}}
\newcommand{\kop}[1]{\;\mathbf{#1}\;}
\newcommand{\kpre}[1]{\mathbf{#1}\;}

\newcommand{\type}[1]{\mathsf{#1}}
\newcommand{\op}[1][op]{\mathsf{#1}}

\newcommand{\seq}[2]{\kpre{do} #1 \leftarrow #2 \kop{in}}
\newcommand{\letseq}[2]{\kpre{let} #1 = #2 \kop{in}}
\newcommand{\conditional}[3]{\kpre{if} #1 \kop{then} #2 \kop{else} #3}
\newcommand{\fls}{\kord{false}}
\newcommand{\fun}[1]{\kpre{fun} #1 \mapsto}

\newcommand{\handler}{\kpre{handler}}
\newcommand{\opgen}[1][\op]{#1}
\newcommand{\opcall}[4][\op]{#1(#2; #3.\,#4)}
\newcommand{\opclause}[3][\op]{#1(#2; #3) \mapsto}
\newcommand{\return}{\kpre{return}}
\newcommand{\tru}{\kord{true}}
\newcommand{\retclause}[1]{\return #1 \mapsto}
\newcommand{\withhandle}[2]{\kpre{with} #1 \kop{handle} #2}

\newcommand{\listtype}{\ \type{list}}
\newcommand{\reftype}{\ \type{ref}}
\newcommand{\unittype}{\type{unit}}

\newcommand{\iref}{{}\mathrel{\kpre{ref}}}
\newcommand{\bang}{!}
\newcommand{\anon}{\mathord{\rule[-0.3pt]{1.2ex}{1pt}}}

\newcommand{\get}{\op[get]}
\newcommand{\set}{\op[set]}

\newcommand{\unt}{\kord{()}}

\newcommand{\step}{\leadsto}

\newcommand{\typar}{\alpha}
\newcommand{\fra}[1]{\forall #1 .}
\newcommand{\boolty}{\type{bool}}
\newcommand{\C}{\underline{C}}
\newcommand{\D}{\underline{D}}
\newcommand{\hto}{\Rightarrow}
\newcommand{\opto}{\to}
\newcommand{\E}{\mathop{!}}

\newcommand{\ctx}{\tyctx; \polyctx; \monoctx}
\newcommand{\shadectx}{\shade{\tyctx; \polyctx;} \monoctx}
\newcommand{\tyctx}{\Theta}
\newcommand{\monoctx}{\Gamma}
\newcommand{\polyctx}{\Xi}
\newcommand{\ent}{\vdash}
\newcommand{\sig}{\Sigma}
\newcommand{\T}{\mathrel{:}}
\newcommand{\union}{\cup}

\newcommand{\figheading}[1]{{\bf\textit{#1}}}
\definecolor{shade}{RGB}{223,223,223}
\usepackage{tcolorbox}
\newtcbox{\shadebox}{on line,arc=1pt, outer arc=2pt,%
  colback=shade,colframe=shade,boxsep=0pt,%
  left=1pt,right=1pt,top=2pt,bottom=2pt,%
  boxrule=0pt,bottomrule=1pt,toprule=1pt}
\newcommand{\shade}[1]{%
        \shadebox{\ensuremath{#1}}%
}

\newtcbox{\myfigbox}{%
  on line, arc=1pt, outer arc=1pt,colframe=black,
  colback=white,
  boxsep=0pt,
  left=1pt,right=1pt,top=1pt,bottom=1pt,%
  boxrule=.5pt,bottomrule=.5pt,toprule=.5pt}

\newtcbox{\myfigboxshade}{%
  on line, arc=1pt, outer arc=1pt,colframe=black,
  colback=shade,boxsep=0pt,
  left=1pt,right=1pt,top=1pt,bottom=1pt,%
  boxrule=.5pt,bottomrule=.5pt,toprule=.5pt}

\newcommand{\figsep}{\hrulefill\par\medskip}

\newcommand{\lst}{\vec}
\usepackage{suffix}
\WithSuffix\newcommand\lst*{\overrightarrow}
\newcommand{\length}[1]{\left\lvert#1\right\rvert}

\newcommand{\doop}{(\textsc{do-op})}
\WithSuffix\newcommand\doop*{\textsc{do-op}}
\newcommand{\withhandled}{(\textsc{handled-op})}
\WithSuffix\newcommand\withhandled*{\textsc{handled-op}}
\newcommand{\withunhandled}{(\textsc{unhandled-op})}
\WithSuffix\newcommand\withunhandled*{\textsc{unhandled-op}}
\newcommand{\gen}{(\textsc{gen})}
\WithSuffix\newcommand\gen*{\textsc{gen}}

\newcommand{\runst}{H_{ST}}
\newcommand{\impmap}{\mathrm{imp\_map}}
\newcommand{\makeref}{\mathrm{make\_ref}}
\newcommand{\reverse}{\mathrm{reverse}}
\newcommand{\foldl}{\kpre{foldl}}
\newcommand{\unitval}{\unt}
\newcommand{\cons}{\mathrel{::}}
\newcommand{\effcomment}[1]{(*\ \text{#1}\ *)}
\newcommand{\nil}{[\,]}
\newcommand{\id}{\mathrm{id}}
\newcommand{\listid}{\mathrm{list\_id}}
\newcommand{\nilv}{\mathrm{nil}}

\newcommand{\?}{;}
\newcommand{\statesim}[2]{\langle #1, #2 \rangle}

\newcommand\inbnfor{\mathrel{\big|}}
\newcommand\mpu{\mathtt{p}}
\newcommand\mpv{\mathtt{q}}
\newcommand\mpw{\mathtt{r}}
\newcommand\deref[1]{!#1}
\newcommand\assign[2]{#1 := #2}
\newcommand\dlet[2]{\kpre{dlet} #1 \leftarrow #2 \kop{in}}
\newcommand\hole{[\ ]}
\newcommand\boundparams[1]{\mathrm{bp}(#1)}
\usepackage{mathtools}
\DeclarePairedDelimiter{\trans}{\lceil}{\rceil}
\DeclarePairedDelimiter{\ctrans}{\lfloor}{\rfloor}
\newcommand\dstep{\mathrel{\overset{dyn}\longleadsto}}
\WithSuffix\newcommand\dstep+{\mathrel{\joinrel{\overset{dyn}{%
        \longleadsto
}}\!\!{}^+}}

\newcommand\sstep{\mathrel{\overset{st}\leadsto}}
\WithSuffix\newcommand\sstep+{\mathrel{\joinrel{\overset{st}{%
        \leadsto
}}\!\!{}^+}}

\WithSuffix\newcommand\step+{\mathrel{\leadsto\!\!\!\!^+}}
\WithSuffix\newcommand\step*{\mathrel{\leadsto\!\!\!\!^*}}

\newcommand\dent[1][\sig]{\vdash^{\mathrm{dyn}}_{#1}}

\usepackage{okexplain}

\title[No value restriction is needed for algebraic effects and handlers]%
      {No value restriction is needed\\for algebraic effects and
  handlers}
\author[O.~Kammar and M.~Pretnar]{%
  Ohad Kammar\thanks{Supported by the European Research Council grant
    `events causality and symmetry --- the next-generation semantics',
    and the Engineering and Physical Sciences Research Council grant
    `quantum computing as a programming language'.}\textdagger\\
  University of Cambridge Computer Laboratory and\\
  University of Oxford Department of Computer Science, England
 \\and\\
 Matija Pretnar\thanks{The material is based upon work supported by
    the Air Force Office of Scientific Research, Air Force Materiel
    Command, USAF under Award No. FA9550-14-1-0096.}\\
 Faculty of Mathematics and Physics, University of Ljubljana,
 Slovenia
}%

\makeatletter

\patchcmd{\NAT@test}{\else \NAT@nm}{\else \NAT@nmfmt{\NAT@nm}}{}{}

\DeclareRobustCommand\citepos
  {\begingroup
   \let\NAT@nmfmt\NAT@posfmt
   \NAT@swafalse\let\NAT@ctype\z@\NAT@partrue
   \@ifstar{\NAT@fulltrue\NAT@citetp}{\NAT@fullfalse\NAT@citetp}}

\let\NAT@orig@nmfmt\NAT@nmfmt
\def\NAT@posfmt#1{\NAT@orig@nmfmt{#1's}}

\makeatother
\begin{document}
\maketitle

\begin{abstract}
  We present a straightforward, sound Hindley-Milner polymorphic type
  system for algebraic effects and handlers in a call-by-value
  calculus, which allows type variable generalisation of arbitrary
  computations, not just values.  This result is surprising.  On the
  one hand, the soundness of unrestricted call-by-value Hindley-Milner
  polymorphism is known to fail in the presence of computational
  effects such as reference cells and continuations.  On the other
  hand, many programming examples can be recast to use effect handlers
  instead of these effects.  Analysing the expressive power of effect
  handlers with respect to state effects, we claim handlers cannot
  express reference cells, and show they can simulate dynamically
  scoped state.
\end{abstract}

\section{Introduction}
\label{sec:introduction}

The following \emph{OCaml}
example~\citep{garrigue-relaxing-the-value-restriction} demonstrates
the problematic interaction between Hindley-Milner polymorphism, which
increases code reuse, and computational effects, such as reference
cells, in a call-by-value language:
\begin{align*}
&\letseq{r}{\iref\nil} && \text{$(*$ generalise $r : \forall \alpha . \alpha\listtype\reftype$ $*)$}\\
&r := [\unitval];        && \text{$(*$ specialise $\alpha := \unittype$ $*)$} \\
&\tru :: \, \bang r      && \text{$(*$ specialise $\alpha := \boolty$ $*)$}
\end{align*}
A na\"\i ve type inference algorithm would assign the type
$\alpha\listtype\reftype$ to the term $\iref\nil$. Unrestricted, it
would assign to $r$ the \emph{type scheme}
$\forall \alpha . \alpha\listtype\reftype$. But doing so allows us to
instantiate $r$ with the unit type $\alpha := \unittype$ to store the
singleton list with the unit value, and then to instantiate $r$ with
the boolean type $\alpha := \boolty$. The result is a list whose
second element is the unit value, but appears to the type system as a
list of booleans.

The current way to avoid this well-known unsound
behaviour~\citep{pierce:taplb,DBLP:journals/lisp/HarperL93,remy:types-lecture-notes}
is to enforce a \emph{value restriction}: the inference algorithm will
generalise the type variables only in value terms that cannot be
reduced further~\citep{wright-simple-imperative-polymorphism}. While
this restriction can be weakened to allow some
computation~\citep{garrigue-relaxing-the-value-restriction}, it still
rules out sound pure programs:
\begin{align*}
&\letseq{id}{(\fun f f)\ (\fun x x)}  & \text{$(*\ \mathrm{id}$ is \emph{not} polymorphic $*)$}\\
&id\,(id)
\end{align*}

The problem only arises when all three components are present:
computational effects, polymorphism, and call-by-value. Without
effects, Milner's original calculus soundly integrates call-by-value
with type inference \citep{MILNER1978348}. Without polymorphism,
computational effects behave predictably in call-by-value
languages. Without call-by-value,
\cite{leroy-polymorphism-by-name-for-references-and-continuations}
combines computational effects with polymorphism without
restriction. \citeauthor{leroy-polymorphism-by-name-for-references-and-continuations}'s
language has two constructs for sequencing: a call-by-name polymorphic
$\letseq x{c_1}{c_2}$ construct in which $c_1$ is re-executed whenever
it is specialised in $c_2$, and a call-by-value monomorphic
$\seq x{c_1}{c_2}$ construct in which $c_1$ is only evaluated once,
but its type is not generalised. The situation is identical in the
\emph{Haskell} programming language, from which we borrowed this
notation.

Programming with algebraic effects and
handlers~\citep{bauer-pretnar:programming-with-algebraic-effects-and-handlers}
is a new approach to structuring functional programs with
computational effects.  The programmer declares a collection of
\emph{algebraic effect operations} with which she structures her
effectful code. Then, separately, she defines \emph{effect handlers}
that implement these abstract operations.
\citeauthor{bauer-pretnar:programming-with-algebraic-effects-and-handlers}'s
\Eff{} programming language is a strict (i.e., call-by-value)
functional language with Hindley-Milner polymorphism, in which all
computational effects are treated as algebraic effects that can be
handled. There is a pre-defined collection of effects that receive
special treatment: runtime errors and memory accesses.  If these
effects are not handled by the program, the runtime will handle them,
invoking the corresponding real computational effects. As \Eff{}
combines the three problematic components (strictness, polymorphism,
effects), it currently imposes the standard value restriction on the
programmer.

In this paper, we show that if only algebraic effects and handlers are
present, the language does not need a value restriction. We present a
straightforward sound Hindley-Milner polymorphic type system for a
call-by-value language that incorporates computational effects in the
form of algebraic effects and their handlers. In order to simplify the
presentation, we present a type system without its associated complete
inference algorithm. Doing so decouples the algorithmic concerns of
finding principal types and complexity from the semantic concern for
soundness. As first-class polymorphism typically makes type inference
undecidable~\citep{wells:typability-and-type-checking-in-System-F-are-equivalent-and-undecidable},
our type system uses ML-style polymorphism.

The rest of the paper is structured as follows. In
Sec.~\ref{sec:tutorial}, we give a short recap of handlers and show
how they may be used to simulate global state. Next, in
Sec.~\ref{sec:type-system}, we give a type and effect system and
sketch the proof of its soundness. We formalized the proof in the
Twelf proof assistant~\citep{twelf}, extending
\citepos{bauer-pretnar:an-effect-system-for-algebraic-effects-and-handlers}
existing formalization of \Eff{}'s core calculus. In
Sec.~\ref{sec:discussion} we evaluate our type system and discuss its
expressivity with respect to mutable references and dynamically scoped
state. Sec.~\ref{sec:conclusion} concludes.

\section{Handlers of algebraic effects}
\label{sec:tutorial}

\emph{Algebraic effects} are an approach to computational effects based on a
premise that impure behaviour arises from a set of \emph{operations}
such as $\op[get]$ and $\op[set]$ for mutable store, $\op[read]$ and
$\op[print]$ for interactive input and output, or $\op[raise]$ for
exceptions~\citep{DBLP:journals/acs/PlotkinP03}.
This naturally gives rise to \emph{handlers} not only of
exceptions, but of any other effect, yielding a novel concept that, amongst
others, can capture backtracking, co-operative
multi-threading, Unix-style stream redirection, and delimited
continuations~\citep{plotkin-pretnar:handling-algebraic-effects,bauer-pretnar:programming-with-algebraic-effects-and-handlers}.

\subsection{Language}

We base our development on the calculus (Fig.~\ref{fig:syntax}) given
in \citepos{pretnar:tutorial} tutorial.
\begin{figure}
\figsep
\input{syntax}
\figsep
\input{sugar}
\figsep
\caption{an idealised calculus of effect handlers}
\label{fig:syntax}
\end{figure}
The language is a variant of the fine-grained call-by-value $\lambda$-calculus
of \cite{levy-power-thielecke:modelling-environments-in-call-by-value-programming-languages},
in which terms are split into inert \emph{values} and potentially effectful
\emph{computations}.

Programmers introduce effects with the construct $\opcall{v}{y}{c}$,
which calls the operation $\op$ with the parameter $v$. The effect
invocation may yield a value to the continuation $c$ using the bound
variable $y$. Programmers define the meaning of such operation calls by
enclosing them in effect handlers.  A handler specifies a return
clause, used when the computation returns a final value, and a
collection of operation clauses $\opclause xkc$, which specify how we
should execute an invocation of the operation $\op$ called with the
parameter $x$ and a continuation $k$. The underlying idea is that
operation calls behave as signals that propagate outwards until they
reach a handler with a matching clause.

Our handlers are \emph{deep}: the additional effects in the
continuation are also handled by the current handler. Our handlers are
also \emph{forwarding}: unhandled operations propagate through each
handler until they are handled or reach the top level. None of these
design choices is essential to the development below, but we make them
to mirror \Eff{}'s design choices.

We use the following syntactic sugar (Fig.~\ref{fig:syntax}):
semicolons elaborate to binding fresh (dummy) variables; function
calls, conditionals, and operation calls are elaborated to
call-by-value evaluation order; function introduction may abstract
over multiple arguments; and bare operations without a parameter or a
continuation argument elaborate to the corresponding \emph{generic
  effect}~\citep{DBLP:journals/acs/PlotkinP03}. In our examples, we
further assume to have the type $\unittype$ with the sole inhabitant
$\unitval$.

\subsection{State handlers}
\label{subsec:state handlers}
We represent state with an operation $\set$, which sets the state
contents to a given parameter and returns $\unitval$, and $\get$,
which takes a unit parameter and returns the state contents. For
example, here is a computation that toggles the state and returns the
old value:
\begin{align*}
  T \defeq
  &\conditional{\get \  \unitval}{\\
  &\quad \set \  \fls; \return \tru \\&\!}{\\
  &\quad \set \  \tru; \return \fls}
\end{align*}
As mentioned above, the runtime of \Eff{}~\citep{bauer-pretnar:programming-with-algebraic-effects-and-handlers}
deals with unhandled primitive effects, but in our calculus, the
behaviour of operations will be determined exclusively by handlers,
and the computation $T$ gets stuck when evaluated.

A simple example of a
handler that can handle a stateful computation is one that sets the
state to a fixed value, say $\tru$, and ignores all its modifications:
\[
  H_C := \handler \{\begin{array}[t]{@{}l@{}r*1{@{}l}}
                      \opclause[\get]{&\anon}{k} {}&k \  \tru \\
                      \opclause[\set]{&s}{k} {}&k \  \unt \\
                      \mathrlap{\retclause{x\,}{}}&& \return x
                                    \}\end{array}
\]
Whenever a $\get$ operation is called, we yield $\tru$ to the
continuation, whereas all $\set$ calls are silently ignored by
yielding the expected unit value $\unitval$ and doing nothing
else. The return clause of a handler states that the returned values
are kept unmodified. When we handle $T$ with $H_C$, we get back the
result $\tru$, no matter how many times we call $T$.

A more useful handler is one that handles $\get$ and $\set$ in a way
that results in the expected stateful behaviour. It uses
a technique called
\emph{parameter-passing}~\citep{plotkin-pretnar:handling-algebraic-effects},
where we transform the handled computation into a function that passes
around a parameter, in our case the state contents:
\[
\runst := \begin{aligned}[t]
  &\handler\{\begin{aligned}[t]
\opclause[\!\!\get\,]{\_}{k} &\ \return (\fun s (k\ s) \ s)\\
\opclause[\set]{s'}{k} &\ \return (\fun \_ (k\ \unitval)\ s')\\
 \retclause {x\,}&\ \return (\fun \_ \return x)
  \}
 \end{aligned}
\end{aligned}
\]

We handle $\get$ with a function that takes the current state contents
$s$ and in the first application, passes them as a result of $\get$ to
the continuation.  As our handlers are deep, the continuation is
further handled into a function, which we again need to supply with
the state contents. Since reading does not modify the state, we again
pass $s$. We handle $\set$ by first passing the unit result, and then
applying the handled continuation to the new state $s'$ as given by
the parameter of $\set$.  The return clause of $\runst$ also needs to
produce a function that depends on the given state, in particular, a
function that returns the given value regardless of the state
contents.

\subsection{Operational semantics}
\label{sec:operational-semantics}

To see how exactly $\runst$ can be used to simulate state, consider
the operational semantics of the calculus, also copied verbatim
from~\citepos{pretnar:tutorial} tutorial. The semantics is given in
terms of the small-step relation $c \step c'$, defined in
Fig.~\ref{fig:semantics}. As expected, there is no such relation
for values, as these are inert.
\begin{figure}
\figsep
\input{semantics}
\figsep
\caption{operational semantics}
\label{fig:semantics}
\end{figure}

The rules for conditionals and function application are standard. For
the sequencing construct, ${\seq{x}{c_1}{c_2}}$, we start by evaluating
$c_1$. If this returns some value~$v$, we bind it to $x$ and
evaluate~$c_2$. But if $c_1$ calls an operation, we propagate the call
outwards and defer further evaluation to the continuation of the call,
for example:
\begin{align*}
  &\seq{x_1}{(\seq{x_2}{\opcall{x}{y}{c_2}} c_1)} c \ \step \\
  &\seq{x_1}{\opcall{x}{y}{\seq{x_2}{c_2} c_1}} c \ \step \\
  &\opcall{x}{y}{\seq{x_1}{(\seq{x_2}{c_2} c_1)} c}
\end{align*}
In our account, we gloss over the standard issues with
capture-avoiding substitution and implicitly assume the appropriate
freshness conditions. For example, in this case, that $y$ is fresh for
$c_1$.

To evaluate $\withhandle{h}{c}$, we start by evaluating $c$. If it
returns a value, we continue by evaluating the return clause of
$h$. If $c$ calls an operation~$\op$, there are two options. If $h$
has a matching clause for $\op$, we start evaluating that, passing in
the parameter and the continuation. Recall that our handlers are deep,
thus the continuation $k$ are also handled by the current handler,
see~$\withhandled*$. If $h$ does not have a matching clause, we
forward the call outwards just like in sequencing,
see~$\withunhandled*$.

Let us return to the state handler $\runst$. If we use it on a
stateful computation, no effects occur as the handled computation
returns a function waiting for an initial state.  To run it, we need
to apply this function to the initial state.  Let us abbreviate such
an application by:
\[
  \statesim{c}{s} := (\withhandle{\runst}{c}) \  s
\]
(note that we use the syntactic sugar for call-by-value function calls
from Fig.~\ref{fig:syntax}).

Even though our calculus is pure, we can show the handler $\runst$
simulates global state in the following way. Let $\sstep$ be the usual
small-step semantics for global state, i.e.:
\begin{mathpar}
  \statesim{\opgen[\get]{\unitval}}s \sstep \statesim{\return s}s

  \statesim{\opgen[\set]{(s')}}s \sstep \statesim{\return \unitval}{s'}

  \inferrule{
    \statesim{c_1}{s} \sstep \statesim{c'_1}{s'}
  }{
    \statesim{\seq x{c_1}{c_2}}{s} \sstep \statesim{\seq x{c'_1}{c_2}}{s'}
  }
\end{mathpar}
etc.

We can prove that for each
$\statesim{c_1}{s} \sstep \statesim{c'_1}{s'}$, we have
$\statesim{c_1}{s} \step+ \statesim{c'_1}{s'}$, and therefore effect
handlers simulate the operational semantics for global state.  For
example:
\begin{align*}
  \statesim{\opgen[\get]{\unitval}}{s}
  & \step
  (\withhandle{\runst}{(\opcall[\get]{\unitval}{y}{\return y})}) \  s \\
  & \step
  (\fun{s'} ((\fun y \withhandle{\runst}{(\return y)}) \ s') \ s') \  s \\
  & \step
  ((\fun y \withhandle{\runst}{(\return y)}) \ s) \ s \\
  & \step
  (\withhandle{\runst}{(\return s)}) \ s \\
  & =
  \statesim{\return s}{s}
\end{align*}
Similarly, we can prove:
\begin{mathpar}
  \statesim{\opgen[\set]{(s')}}{s} \step+ \statesim{\return \unitval}{s'} \\
\end{mathpar}
For the third transition, case-split on the possible transitions
$\statesim {c_1}s \step \statesim {c'_1}{s'}$.

In summary, the $\runst$ handler faithfully simulates state. For more
details on simulating state,
see~\cite{bauer-pretnar:an-effect-system-for-algebraic-effects-and-handlers}
and \cite{danvy-advanced-thesis}. Therefore, even though our calculus
is pure, it faithfully simulates impure computation. By giving an
unrestricted Hindley-Milner type system to this calculus, we now
show that the effects expressible by effect handlers interact well
with polymorphism.

\section{Type system}
\label{sec:type-system}

\begin{figure}
\figsep{}
\input{types}
\figsep{}
\caption{types and effects}
\label{fig:types}
\end{figure}
\begin{figure}
\figsep
\input{type-system-rules}
\figsep
\caption{a polymorphic type and effect system}
\label{fig:type-system}
\end{figure}

The type and effect system
(Figs.~\ref{fig:types}--\ref{fig:type-system}) closely
follows~\cite{pretnar:tutorial}. It comprises two kinds of types:
values are typed with simple types $A$, while the types of
computations are additionally annotated with finite sets of operations
$\sig$ like in an effect system of~\cite{lucassen-gifford:polymorphic-effect-systems}.

We modify Pretnar's system in two ways. The first modification is
minor. We generalise the type system to allow for more flexible
\emph{local} operation signatures $\sig$, where operations may have
different types when handled by different handlers, as in
\cite{kammar-lindley-oury:handlers-in-action}. In contrast,
\citeauthor{pretnar:tutorial}'s account posits a global assignment of
predefined types to the effect operations, and the effect annotations
$\sig$ only list which operations may be present. Local signatures
allow the same operation symbol to appear in disjoint parts of the
program with different types. Local signatures also give the calculus
stronger theoretical properties, such as strong normalisation and
simpler denotational semantics,
cf. \citeauthor{kammar-lindley-oury:handlers-in-action}.

The second modification is our main contribution. We incorporate
Hindley-Milner polymorphism in a standard way, without any value
restriction. We indicate these latter modifications by \shade{shading}
in the figures. Amongst these:
\begin{itemize}
\item Local effect signatures $\sig$ are finite mappings from
  operations $\op$ to pairs of value types $A$, $B$, whose action we
  denote by $(\op : A \to B) \in \sig$. We denote the restriction of a
  signature $\sig$ to the set of operations disjoint from a given set
  $\Delta = \{\op_i \mid 1 \leq i \leq n\}$ by
  $\sig \setminus \Delta$.
\item We extend types with \emph{type variables}~$\typar$ and add
  \emph{type variable environments}~$\tyctx$, which are just finite
  sets of type variables.
\item We introduce \emph{schemes}~$\fra{\lst\typar} A$, where
  $\lst\typar$ denotes a finite set of $\length{\lst\typar}$-many type
  variables ranged over by $\typar_i$.
\item We introduce \emph{kinding judgements} $\tyctx \ent X$ to
  explicitly keep track of the free type variables in $X$. The
  shorthand $\tyctx \ent X, Y, Z$ stands for the conjunction $\tyctx
  \ent X$, $\tyctx \ent Y$, and $\tyctx \ent Z$.
\item Typing judgements $\ctx \ent M : X$ include the standard
  monomorphic environments $\monoctx$ which are a unique assignment of
  types to variables. We extend those with type variable
  environments~$\tyctx$ and \emph{polymorphic environments}~$\polyctx$,
  which are a unique assignment of schemes to variables. We assume
  that no variable can appear in both $\monoctx$ and
  $\polyctx$.\footnote{%
      This separation into two environments is not strictly necessary, as
      a monomorphic environment $\monoctx$ may be identified with a
      polymorphic environment where each quantifier ranges over an empty
      tuple of type variables. We choose to separate the two to
      highlight which parts of the language interact with
      polymorphism.
    } These polymorphic variables can be specialised at any type.
  \item We add \emph{scheme judgements} whose effect annotation is
    outside the scope of the quantifier. The kinding assumption
    $\tyctx \ent\sig$ ensures that none of the type variables
    $\lst\typar$ appears in $\sig$. It is at this point that the
    decision of the inference algorithm which type variables
    $\lst\typar$ to generalise over takes effect. Our choice to
    separate scheme judgements from type judgements simplifies the
    let-rule, and makes it very similar to its standard, monomorphic
    counterpart.
\end{itemize}

The remaining kinding and typing rules are standard. Fine-grained
call-by-value functions take values and perform computations. An
operation invocation is well-typed if the type assigned to it by the
local signature must agree with the type of the given parameter value
$v$, and with the type of argument the continuation $c$ expects. A
handler is well-typed if the type of result the return clause expects
matches with the type of computation the handler can handle, and each
operation clause is well-typed when the parameter type and
continuation type match the local signature the handler can
handle. Both clauses can cause additional effects, and their effect
annotation must include these operations, as well as any effect
operations the handler does not explicitly handle, reflecting the fact
that our handlers are forwarding. Thus, the rule also requires the
type and effect of both clauses to agree. The fact that our handlers
are deep is reflected by the type of the continuation: the effects the
continuation may cause have already been handled, and so the
continuation may cause effects in the resulting signature and of the
resulting return type.

For the given effect system, we then have:

\begin{theorem*}[Safety]
If\/ $\ent c \T A \E \sig$ holds, then either:
\begin{enumerate}[(i)]
\item
    $c \step c'$ for some $\ent c' \T A \E \sig$;
\item
    $c = \return v$ for some $\ent v \T A$; or
\item\label{op-clause}
  $c = \opcall{v}{y}{c'}$ for some $(\op \T A_{\op} \to B_{\op}) \in \sig$, $\ent
  v \T A_{\op}$, and $y \T B_{\op} \ent c' \T A \E \sig$.
\end{enumerate}
\end{theorem*}
In particular, when $\sig = \emptyset$, evaluation will not get stuck
before returning a value.
\begin{proof}
  We prove progress and preservation lemmata separately by
  induction. We formalized\footnote{%
    \url{https://github.com/matijapretnar/twelf-eff/tree/val-restriction-local-sig}} %
  the calculus and the safety theorem in the Twelf proof
  assistant~\citep{twelf}. Our formalization extends
  \citepos{bauer-pretnar:an-effect-system-for-algebraic-effects-and-handlers} existing formalization
  of Eff's core calculus with type schemes and polymorphism. The code
  is compatible with version~1.7.1 of Twelf. We summarise the crucial
  step, namely proving type and effect preservation under the $\doop*$
  transition.

Assume that the reduct in $\doop*$ is well-typed, and invert its type derivation:
\begin{mathpar}
\inferrule{
  \inferrule{
  \inferrule
  {\inferrule{~}{(\op \T A_{\op} \opto B_{\op}) \in \sig} \
    \inferrule{\vdots}{\tyctx,\lst\typar\ent v \T A_{\op}} \
    \inferrule{\vdots}{\tyctx,\lst\typar;y \T B_{\op} \ent c_1 \T A \E \sig}}
  {\tyctx,\lst\typar\vdash \opcall{v}{y}{c_1} :  A!\sig}}
  {\tyctx \vdash \opcall{v}{y}{c_1} : (\forall\lst\typar. A)\E\sig}
    \  \inferrule{\vdots}{\tyctx;x : \forall\lst\typar. A\vdash c_2 : B \E \sig}
}{
  \tyctx\vdash \seq{x}{\opcall{v}{y}{c_1}} c_2 : B \E \sig
}
\end{mathpar}

The $\gen*$ rule ensures that none of the type variables in
$\lst\typar$ appear in $\sig$. Because $\sig$ includes
$\op \T A_{\op} \opto B_{\op}$, none of these variables appear in
$A_{\op}$, and we may strengthen the derivation of
$\tyctx,\lst\typar\ent v \T A_{\op}$ to a derivation of
$\tyctx\ent v \T A_{\op}$. As a consequence, the following
derivation is valid:
\begin{mathpar}
\inferrule{
  \inferrule{~}{(\op \T A_{\op} \opto B_{\op}) \in \sig}\
  \inferrule{\vdots}{\tyctx \ent v \T A_{\op}}\
  \inferrule{
    \inferrule{\inferrule{\vdots}{\tyctx,\lst\typar;y \T B_{\op} \ent c_1 \T A \E \sig}}
              {\tyctx;y \T B_{\op} \ent c_1 \T (\forall\lst\typar.A) \E \sig}\
    \inferrule{\vdots}{\tyctx;x : \forall\lst\typar. A\vdash c_2 : B \E \sig}
  }{
  \tyctx; y \T B_{\op} \ent \seq{x}{c_1} c_2 \T B \E \sig}}
{
  \tyctx\vdash\opcall{v}{y}{\seq{x}{c_1} c_2} : B \E \sig
}
\end{mathpar}
Therefore, the reduction in $\doop*$ preserves both the type and the
effect annotation.
\end{proof}
The Safety Theorem is robust under the following standard variations
in the calculus:
\begin{description}
\item[coarse annotations.]%
  \label{localise}%
  We can make the signature $\sig$ global, and only keep track of
  which operations are used, as in~\cite{pretnar:tutorial}. The types
  in this global signature cannot use any type variables. The
  soundness proof remains essentially
  unchanged\footnote{\url{https://github.com/matijapretnar/twelf-eff/tree/val-restriction-global-sig}}.
  Due to the lack of type variables in the global signature, there is
  no need to impose a side-condition on the well-formedness of the
  effect annotation in the $\gen*$ rule

  It may seem this coarser system is a restriction of our current
  system, where the type information for each operation has to agree
  in all effect annotations, and hence it is sound by the Safety
  Theorem. This is not the case. In this coarser system, the
  signatures on function types are not annotated with the types of the
  operations. If those types were fully written out, they would
  involve the global signature, leading to potential mutual recursion
  between signatures and function types. For example, if we elaborate
  the global signature
  $\sig = \{ \op : \unittype \to (\unittype \to \unittype)\}$, we
  would get:
  \[
  \sig = \{ \op : \unittype \to (\unittype \to (\unittype \E \sig)) \}
  \]
  where left arrow is part of the signature syntax and receives no effect
  annotation on the co-domain.
  This recursion is not a mere formality. The type-and-effect system
  with local signatures we have described ensures well-typed terms
  terminate, cf.~\cite{kammar-lindley-oury:handlers-in-action}. When
  we switch to a global signature, we can use effect operations with
  higher-order return types to express well-typed diverging
  computations. With the above global signature
  $\sig = \{ \op : \unittype \to (\unittype \to \unittype)\}$,
  consider the handler
  \[
    H := \handler \{\begin{array}[t]{*3{@{}l}}
                      &\retclause x\return x,\\
                      &\opclause \_k k(\fun \_\opgen\,
                        \unitval\,\unitval)\}\end{array}
  \]
  In the coarse type system, we can derive the judgement:
  \[
    \ent H : (\unittype \E \{\op\}) \hto (\unittype \E \emptyset)
  \]
  If we handle the simple looking computation
  $\ent \op\,\unitval\,\unitval \T \unittype \E \{\op\}$ with
  $H$, we get a diverging computation:
  \begin{align*}
    \withhandle H
    {\opgen\,\unitval\,\unitval}
    &\step+
    \withhandle H{\big(\fun \_\opgen\,\unitval\,\unitval\big)\,\unitval}
    \\&\step\ \,
    \withhandle H{\opgen\,\unitval\,\unitval}
  \end{align*}
  In fact, by a variation on
  \citepos{landin:the-mechanical-evaluation-of-expressions} knot, we
  can express a variant of the $Y$-combinator,
  such that for a function $f$ that is pure,
  $Y\,f$ behaves like the fixed-point of $f$ when invoked on pure arguments.

\item[no annotations.]%
  \label{discard}%
  We can remove all the effect annotations~$\sig$ from type judgements
  and fix a single, global signature $\sig$.  The advantage of having
  an effect system is the additional guarantee in
  clause~\emph{(\ref{op-clause})} of the Safety Theorem, which ensures
  that any unhandled operation must appear in $\sig$. Without
  annotations, any operation may be called. This system is a
  restriction of the coarse variation, where each effect annotation is
  the entire signature. Consequently, it is sound.

\item[additional language features.]%
  \label{orthogonal}%
  To the calculus with coarse annotations, we can add \emph{structural
    subtyping} and static \emph{effect instances} (further discussed
  in Sec.~\ref{subsec:reference-cells}).  The soundness proof remains
  essentially unchanged\footnote{%
    \url{https://github.com/matijapretnar/twelf-eff/tree/val-restriction-instances}%
  } as these modifications are orthogonal to polymorphism.  Similarly,
  we can replace deep handlers with shallow ones, as in
  \cite{kammar-lindley-oury:handlers-in-action} and
  \cite{kiselyov-sabry-swords:extensible-effects}.  As the
  changes\footnote{%
    \url{https://github.com/matijapretnar/twelf-eff/tree/val-restriction-shallow}%
  } are again orthogonal to polymorphism, we may reasonably assume a
  similar soundness result to hold for a calculus that incorporates
  all of the above: subtyping, instances, and, through two separate
  syntactic constructs, both deep and shallow handlers.
\end{description}
\vspace{-\baselineskip}%
\noindent

\section{Expressivity}
\label{sec:discussion}
There is currently no simple type system integrating reference cells
with polymorphism without the value restriction. This non-existence
contrasts the simplicity of our type system, and calls into question
both its degree of feature integration and its expressiveness. First,
we evaluate the degree and smoothness of the interaction between
polymorphism and other features in our calculus. Then, we highlight
the difference in expressiveness between effect handlers and reference
cells. As a basis for our evaluation and comparison, we use
\citepos{leroy-thesis} set of example programs for analysing the
usefulness of a polymorphic type system for reference cells.

\subsection{Evaluation}
\label{subsec:evaluation}
Algebraic effects allow us to lace a piece of code with operations in the
signature \[ \{\get \T\unittype\to\alpha,
\set\T\alpha\to\unittype\} \] The scheme assigned to the handler $\runst$,
which handles them away, is
\[
\runst \T \forall\alpha,\beta.\alpha\E\{\get \T\unittype\to\beta,
\set\T\beta\to\unittype\}\hto(\beta \to \alpha\E\emptyset)\E\emptyset
\]
It takes a computation of type $\alpha$ that interacts with a state of type $\beta$,
and handles it to a pure function of type $\beta \to \alpha \E \emptyset$.
The rightmost $\emptyset$ indicates that no effects can occur when producing the function.

This handler can handle computations with different types of state,
for example:
\begin{align*}
  &(\withhandle {\runst} {\set \ \unitval})\ \unitval;\\
  &(\withhandle {\runst} {\get \ \unitval})\ \tru
\end{align*}
We can also use effects in polymorphic code:
\begin{align*}
&\seq f{
\begin{aligned}[t]
  \conditional{\get \ \unitval}{&\return\fun {x\,y} \return x\\}
                        {&\return \fun {x\,y} \return y}
\end{aligned}\\&}
\begin{aligned}[t]
(f\ &(\fun b \return b)\\
    &(\fun b \set\ b;\return b))\\
  (&f\ \tru\ \fls)
\end{aligned}
\end{align*}
In our call-by-value semantics, if we wrap this computation with the
state handler, the memory look-up in $f$'s definition will only occur
once.

To demonstrate that the polymorphic, effectful, and high-order features
interact well, we hypothetically extend our calculus with pairs and
lists. The hypothesised extension may include primitives such as the
empty list $\nil$, a list cons $(\cons)$ and tail-recursive iteration
$\kord{foldl}$, which we expect to interact smoothly with
polymorphism.  Thus we can use $\runst$ to implement functional
features in an imperative style.
\[
\begin{array}{*{3}{@{}l}}
\seq{\ &\impmap}{\fun {f\,xs}\\
& \withhandle{\runst}{
  (\begin{array}[t]{@{}l}
  \foldl\ \begin{aligned}[t]&(\fun {x} \set\,(f\ x \cons \get \ \unitval)\\&\unitval\\&xs\?\end{aligned}\\
  \reverse\,(\get \ \unitval)
  \end{array}\\
& \nil\ \effcomment{initial state}
}}\ldots
\end{array}
\] The scheme
assigned to $\impmap$ is
\[
\impmap \T \forall \alpha\beta.(\alpha \to \beta\E\sig)\to(\alpha\listtype \to \beta\listtype \E\sig)\E\emptyset
\]
for any $\sig$.
This implementation is imperative in style, but not imperative
\emph{per se}, as all operations are handled by high-order functions.
The function $\impmap$ can also be partially applied and retain its
polymorphism, for example, in
\[
\begin{array}{*{3}{@{}l}}
&\seq{\listid&{}}{\impmap\ \id}\\
&\seq{\nilv  &{}}{\listid\ \nil}\ldots
\end{array}
\]
we have the scheme assignments:
\[
\begin{array}{*{3}{@{}l}}
\listid &{}\T \forall \alpha . \alpha\listtype \to \alpha\listtype\E\emptyset\\
\nilv   &{}\T\forall \alpha.\alpha\listtype
\end{array}
\]

Most importantly, the following program is well-typed:
\begin{align*}
&\seq{id}{(\fun f f)\ (\fun x x)}\\
&\seq{id'}{id\,(id)}{\ldots}
\end{align*}
and both functions are assigned the polymorphic type
$\forall \alpha. \alpha \to \alpha ! \sig$. Such mixed-variance
polymorphism is ruled out by all current value restrictions.

\subsection{Reference cells}
\label{subsec:reference-cells}
We believe it is impossible to implement full blown reference cells
using effect handlers without other language features. We can increase
modularity by introducing instances~\citep{%
  bauer-pretnar:programming-with-algebraic-effects-and-handlers,
  bauer-pretnar:an-effect-system-for-algebraic-effects-and-handlers,
  pretnar:inferring-algebraic-effects
}.
These may be thought of as first class atomic names. With instances,
each effect instance $\iota$ and an operation symbol $\op$ determine an
operation $\iota \# \op$. In handlers, each operation clause
$v \# \opclause x k c$ specifies which instance, dynamically given
by the value $v$, of the statically chosen effect operation symbol
$\op$ the handler handles. At runtime, invocations of the same
operation $\op$ but with different instances will not be caught by this
handler and will be forwarded.

Instances allow us to pass a cell around by passing an
instance, but they are still less expressive than having the ability to
allocate arbitrarily many new cells dynamically. For example, we do
not know how to implement even the simplest of \citepos{leroy-thesis}
benchmarks:
\[
\seq{\makeref}{\fun x \iref x}\ldots
\]
We believe it is impossible to encode general references without
additional language features. \Eff{} provides such a mechanism, which
can both generate \emph{fresh} instances and attach them to a stateful
\emph{resource}~\citep{%
  bauer-pretnar:programming-with-algebraic-effects-and-handlers
}, allowing one to directly implement a $\makeref$ analogue: $\makeref$
creates a fresh instances that has $\get$ and $\set$ operations
associated with it. Only code that knows what the instance is, can
handle these effects. However, it is not easy to find a corresponding
type and effect system for fresh instances~\citep{%
  bauer-pretnar:an-effect-system-for-algebraic-effects-and-handlers,
  pretnar:inferring-algebraic-effects
}, let alone a polymorphic one.

As a final example, recall the problematic reference cell example
which cannot be directly expressed in our calculus:
\begin{align*}
&\seq{r}{\iref\nil} \\
&r := [\unitval];\\
&\tru :: \, \bang r
\end{align*}
We can express a computation that writes a $\unittype\listtype$ value
and reads a $\boolty\listtype$ value:
\begin{align*}
&\set \ [\unitval];\\
&\tru :: \get \ \unitval
\end{align*}
However, this computation has the effect annotation
\[
\{\set \T
  \unittype\listtype \to \unittype, \get\T \unittype \to \boolty\listtype\}
\]
which is incompatible with the type of the state handler $\runst$.
Other handlers for the state operations may have a compatible type.
For example, the read-only state handler $H_{RO}$ which ignores any
memory updates:
\[
H_{RO} := \begin{aligned}[t]
  &\handler\{\begin{aligned}[t]
 \retclause x&\,\fun \_ \return x\\
\opclause[\get]{\_}{k} &\,\fun s k\ s\ s\\
\opclause[\set]{\_}{k} &\,\fun s k\ \unitval\ s
  \}
 \end{aligned}
\end{aligned}
\]
It has the scheme
\[
H_{RO} \T \forall\alpha,\beta,\gamma.\alpha\E\{\get \T\unittype\to\beta,
\set\T\gamma\to\unittype\}\hto(\beta \to \alpha\E\emptyset)\E\emptyset
\]
and can be applied to the above computation without run-time errors.

\subsection{Dynamically scoped state}

As we saw in Sec.~\ref{subsec:state handlers}, we can simulate
global state using the handler $\runst$, and this state can be handled
locally to give a pure computation. While we do not know whether
effect handlers can simulate reference cells or not, we will now
characterise the handler $\runst$ as expressing the notion of
\emph{dynamically scoped} state.

\begin{figure}
\figsep{}
\input{dynamic-state-syntax}
\figsep{}
\caption{a calculus for dynamically scoped state}
\label{fig:dyn-state-syntax}
\end{figure}

In order to explain what we mean by dynamically scoped state, and to
make the discussion precise, we consider the calculus presented in
Fig.~\ref{fig:dyn-state-syntax}. It is a fine-grained
call-by-value variation on the dynamic scope calculi of
\cite{kiselyov-shan-sabry:delimited-dynamic-binding} and
\cite{moreau:a-syntactic-theory-of-dynamic-binding}.

We assume a set of parameters ranged over by $p$ that name dynamically
scoped memory cells. These cells can be dereferenced, $\deref p$, or
assigned to, $\assign pv$, just like ref cells. The rebinding
construct $\dlet pvc$ declares that in executing $c$, all references
to $p$ will be bound to this occurrence of $p$, and shadow other
binding declarations that may be in place.

For example, assuming we have a type of integers the following code
will evaluate to $\return 2$.\label{dyn-state-example}
\[
  \begin{array}[t]{*3{@{}l}}
  \seq {\,&f}{
           \,\begin{array}[t]{*3{@{}l}}
             \dlet{\,&\mpu} 0\\
          &\return (\fun {\,&\_}\\&&\assign\mpu{1+\deref\mpu})}
           \end{array}\\
  &\begin{array}[t]{*3{@{}l}}
     \dlet{\,&\mpu}1\\
  &f\ ();\\
  &\deref \mpu
   \end{array}
  \end{array}
\]
The reason is that the state changes inside the function bind
dynamically to the closest enclosing rebinding, which is the second
one.

\begin{figure}
\figsep{}
\input{dynamic-state-semantics}
\figsep{}
\caption{semantics for dynamically scoped state}
\label{fig:dyn-state-semantics}
\end{figure}

Fig.~\ref{fig:dyn-state-semantics} describes the (Felleisen-style)
operational semantics for this calculus. We kept the style of
semantics as close as possible to
\citepos{kiselyov-shan-sabry:delimited-dynamic-binding} to make it
clear we use the same notion of dynamic scope, and our theoretical
treatment closely mirrors their own. The semantics use the set of
parameters bound in a given context $E$, denoted by
$\boundparams{E}$. The three transitions specific to dynamic scope are
shaded. First, a fully evaluated computation removes a preceding
parameter binding, as it will no longer be used. For the other two
transitions, the side condition $p \notin \boundparams{E'}$ ensures
the uniqueness of the decomposition into the context $E'$ by locating
the closest rebinding of $p$. The semantics of dereferencing returns
the value associated to this closest rebinding, while the semantics of
assignment modifies it. In our design, assignment evaluates to the
unit value, deviating from
\citeauthor{kiselyov-shan-sabry:delimited-dynamic-binding}'s
semantics. This purely cosmetic change does not alter the nature of
dynamically scope state we are dealing with, and makes the
relationship with $\runst$ tighter.

The example above evaluates as follows:
\[
  \begin{aligned}
  &\begin{array}[c]{*3{@{}l}}
  \seq {\,&f}{
           \,\begin{array}[t]{*3{@{}l}}
             \dlet{\,&\mpu} 0\\
          &\return (\fun {\,&\_}\\&&\assign\mpu{1+\deref\mpu})}
           \end{array}\\
  &\begin{array}[t]{*3{@{}l}}
     \dlet{\,&\mpu}1\\
  &f\ ();\\
  &\deref \mpu
   \end{array}
  \end{array}
             \dstep\quad 
  \begin{array}[c]{*5{@{}l}}
  \seq {\,&f}{\return (\fun {\,&\_}\\&&\assign\mpu{1+\deref\mpu})}\\

  &\begin{array}[t]{*3{@{}l}}
     \dlet{\,&\mpu}1\\
  &f\ ();\\
  &\deref \mpu
   \end{array}
  \end{array}
\\~
\\&           \dstep\quad 
  \begin{array}[c]{*3{@{}l}}
     \dlet{\,&\mpu}1\\
  &(\begin{array}[t]{*3{@{}l}}
           \fun {\,&\_}\\&\assign\mpu{1+\deref\mpu}) \ ();
   \end{array}\\
  &\deref \mpu
   \end{array}
           \quad\dstep\quad 
  \begin{array}[c]{*3{@{}l}}
     \dlet{\,&\mpu}1\\
  &\assign\mpu{1+\deref\mpu};\\
  &\deref \mpu
   \end{array}
           \quad\dstep+\quad 
           \return 2
\end{aligned}
\]

\begin{figure}
\figsep{}
\input{dynamic-state-translation}
\figsep{}
\caption{handlers expressing dynamically scoped state}
\label{fig:dyn-state-translation}
\end{figure}

Fig.~\ref{fig:dyn-state-translation} shows how effect handlers express
dynamically scoped state. Using
\citepos{felleisen:on-the-expressive-power-of-programming-languages}
terminology, it is a \emph{macro} translation. First, it does not use
any information collected globally as it is defined homomorphically
over the syntax of the language. Second, it keeps the common core of
the two languages unchanged, translating a boolean value to itself, a
function to a function, and so forth. The translation is
straightforward: it translates dereferencing and assignments to $p$ as
specially named effects, $\mathtt{get\_}p$ and
$\mathtt{set\_}p$. Rebinding amounts to handling with $\runst$,
and passing the translated rebinding value as the initial value.

This translation simulates dynamic allocation:
\begin{theorem*}[Simulation]
  For all $c \dstep c'$, we have $\trans c \step+ \trans{c'}$.
\end{theorem*}
\begin{proof}
  First, extend the translations to evaluation contexts, and show that
  $\trans{E[c]} = \trans{E}[\trans c]$.  Then, show the translation
  respects capture avoiding substitution:
  $\trans{c[v/x]} = \trans{c}[\trans v/x]$. To deal with the mismatch
  between Felleisen-style and small-step semantics, show that for all
  evaluation contexts $E$, if $c \dstep c'$ then
  $\trans{E}[c] \step+ \trans{E}[c']$. It therefore suffices to prove
  the theorem for each of the transitions in
  Fig.~\ref{fig:dyn-state-semantics} specialised to $E := \hole$.

  For each of the common constructs of the two calculi, the proof is
  immediate, for example:
  \[
    \trans{\seq x{\return v}{c}} =
    \seq x{\return \trans v}{\trans c} \step
    \trans{c}[\trans v/x] = \trans{c[v/x]}
  \]

  The next remaining transition amounts to handling a terminal
  computation:
  \begin{align*}
    \trans{\dlet \mpu v{\return v'}}
    &=
    (\withhandle {\runst^p}{\return \trans {v'}})\ \trans v
    \\
    &\step+
    (\fun \_ \return \trans{v'})\ \trans v
    \step
    \return \trans{v'}
  \end{align*}

  For the final two transition, show that, for all contexts $E$,
  parameters $p \notin \boundparams E$, operations $\op$ that is
  either $\mathtt{get\_}p$ or $\mathtt{set\_}p$, and $x$ fresh for $E$,
  we have:
  \[
    \trans{E}[\opcall vxc] \step*
    \opcall vx{\trans{E}[c]}
  \]
  And finally, calculate:
  \begin{align*}
    \trans{\dlet  p v {E[\deref p]}}
    =&
    (\withhandle {\runst^p}{\trans{E}[\opcall[\mathtt{get\_}p]{()}x{\return x}})\ \trans v
    \\\step*&
    (\withhandle {\runst^p}{\opcall[\mathtt{get\_}p]{()}x{\trans{E}[\return x]}})\ \trans v
    \\\step+&
    (\fun s ((\fun x\withhandle {\runst^p}{\trans{E}[\return x]})\ s) \ s)\ \trans v
    \\\step+&
    \withhandle {\runst^p}{\trans{E}[\return x]}\ \trans v
    \\=&
    \trans{\dlet p v{E[\return v]}}
  \end{align*}
  A similar calculation for assignment completes the proof.
\end{proof}

\begin{figure}
\figsep{}
\input{dynamic-state-types}
\figsep{}
\caption{polymorphic types for dynamically scoped state}
\label{fig:dyn-state-types}
\end{figure}

This translation, while being straightforward, also preserves the type
system. Fig.~\ref{fig:dyn-state-types} presents the types for the
calculus. The only notable feature is that, like
\citeauthor{kiselyov-shan-sabry:delimited-dynamic-binding}, we assume
a global signature assigning to each parameter a type. As the
signature is global, these (monomorphic) types do not contain any type
variables.

\begin{figure}
\figsep{}
\input{dynamic-state-type-system-rules}
\figsep{}
\caption{a polymorphic type system for dynamically scoped state}
\label{fig:dyn-state-type-system-rules}
\end{figure}

Fig.~\ref{fig:dyn-state-type-system-rules} presents the kind and
(Hindley-Milner polymorphic) type system for the calculus. The kind
system ensures well-kinded signatures assign types without type
variables. Typing judgements $\ctx \dent c : A$ refer to the fixed,
ambient, well-kinded parameter signature $\sig$. The typing rules
specific to dynamically scoped state (shaded) ensure that we may only
dereference, assign to, and rebind a parameter in accordance with the
ambient signature. The assignment rule also
highlights our decision to ascribe the unit type to assignment, in a
minor deviation from
\citeauthor{kiselyov-shan-sabry:delimited-dynamic-binding}. The \gen{}
rule is now completely unrestricted, ensured by the assumption that
the type signature does not involve type variables.

Fig.~\ref{fig:dyn-state-type-translation} extends the translation to
types. The parameter signature $\sig$ translates into an effect
signature containing the distinct pair of effects corresponding to
this parameter, namely $\mathtt{get\_}p$ and $\mathtt{set\_}p$, with
the appropriate type. Function types may cause any effect in this
translated signature $\trans\sig$. This translation is therefore
not-well-defined: if $\sig$ contains any function types, then
$\trans\sig$ refers to $\trans{A\to B}$, which refers to $\trans\sig$
again.

\begin{figure}
\figsep{}
\input{dynamic-state-type-translation}
\figsep{}
\caption{handlers types system expressing dynamically scoped state}
\label{fig:dyn-state-type-translation}
\end{figure}

There are at least three ways around this issue. The simplest
solution, presented in the top half of
Fig.~\ref{fig:dyn-state-type-translation} is to restrict $\sig$ to
\emph{ground} types, i.e., prohibit storing functions.

A less restrictive solution is to use the coarser type system for
effect handlers that does not track effect annotations at all, and
define $\ctrans{A \to B} := \ctrans A \to \ctrans B$, as in the bottom
half of Fig.~\ref{fig:dyn-state-type-translation}. This solution works
well, as the effects $\mathtt{get\_}p$ and $\mathtt{put\_}p$ maintain
their type.

A more sophisticated potential solution is to use equi-recursive
effect signatures. At this point in time, such a type-and-effect
system has not been developed, but we do not foresee any serious
obstacles in developing it: its denotational semantics would involve a
recursive domain equation in the same spirit as in
\cite{bauer-pretnar:an-effect-system-for-algebraic-effects-and-handlers}.

The fact that higher-order parameter types merit domain-theoretic
semantics is not surprising, as such parameters allow non-terminating
programs.  We say that a type $A$ is \emph{inhabited} if there exists
a closed value $\dent v : A$.

\begin{proposition*}
  If $\sig$ contains a higher-order type parameter
  $(p : A \to B) \in \sig$ for some inhabited type $A$, then there is
  a term $c$ satisfying:
  \[
    c \dstep+ c
  \]
\end{proposition*}
\begin{proof}
  Let $\dent v : A$ be an inhabitant of $A$, and take:
  \[
    c:= \begin{array}[t]{*2{@{}l}}
          \dlet {&p}{(\fun a{(\deref p)}\,a)}\\
          &(\deref p)\,v
        \end{array}
  \]
  Then:
  \[
    c \dstep
    \begin{array}{*2{@{}l}}
      \dlet {&p}{(\fun a{(\deref p)}\,a)}\\
             &(\fun a{(\deref p)}\,a)\,v
    \end{array}
    \dstep+
    \begin{array}{*2{@{}l}}
      \dlet {&p}{(\fun a{(\deref p)}\,a)}\\
             &(\deref p)\,v
    \end{array}
    = c
  \]
as required.
\end{proof}

Moreover, every parameter $(p : A \to B)$ lets us define a form of a
fixed-point combinator $Y : ((A \to B) \to A \to B) \to (A \to B)$ by
a variant of Landin's knot, provided the functions passed to this
combinator and their arguments do not involve $p$.

The two proposed translations are correct:
\begin{theorem*}[Type Preservation]
  For every $\ctx\dent c : A$ and $\ctx\dent v : A$, we have:
  \begin{itemize}
  \item If $\sig$ is ground, then
    $\trans\tyctx; \trans\polyctx; \trans\monoctx \ent \trans c : \trans A \E {\trans\sig}$
    and $\trans\tyctx; \trans\polyctx; \trans\monoctx \ent \trans v : \trans A$.
  \item $\ctrans\tyctx; \ctrans\polyctx; \ctrans\monoctx \ent \trans c : \ctrans A$ and
    $\ctrans\tyctx; \ctrans\polyctx; \ctrans\monoctx\ent \trans v : \ctrans A$.
  \end{itemize}
\end{theorem*}

\begin{proof}
  For the first part only, first show that if $A$ is ground, then
  $\trans A = A$, and so if $\sig$ is a well-kinded ground signature,
  then $\trans\sig$ is well-defined and well-kinded.

  Then the proofs of both parts follow the same lines. By mutual
  induction on the kinding judgements, show that well-kinded types,
  schemes, and contexts translate into well-kinded types, schemes, and
  contexts, respectively. Then show that both translations respect
  type-level substitution:
  \[ \trans{B[A_i/\alpha_i]_{1 \leq i \leq n}} =
    \trans{B}[\trans{A_i}/\alpha_i]_{1 \leq i \leq n}
  \]
  and similarly for the coarse translation.

  Finally, by mutual induction on typing judgements for values and
  computations, and on scheming judgements, show the hypothesis. We
  mention only the interesting cases.

  For dereferencing a cell $(p : A) \in \sig$, by the translation's
  definition,
  \[
    (\mathtt{get\_}p : \unittype \to \trans A) \in \trans \sig
  \] Use this fact to derive that $\trans{\deref p}$ has the type
  $\trans A$. Use a similar argument for assignment.

  Next, show that for all $(p : A) \in \sig$:
  \[
    \trans\tyctx; \trans\polyctx; \trans\monoctx \ent \runst^p : (B \E
    \trans \sig) \hto ((\trans A \to (B \E \trans\sig)) \E \trans\sig)
  \]
  and use this fact, together with the induction hypotheses, to give a
  valid derivation for $\trans{\dlet pvc}$.
\end{proof}

In summary, the handler $\runst$ expresses dynamically scoped state,
in both terms and types.

\section{Conclusion and further work}
\label{sec:conclusion}
Unexpectedly, Hindley-Milner polymorphism integrates smoothly and
robustly with existing type and effect systems for algebraic effects
and handlers. However, combining reference cell allocation with polymorphism
remains an open problem, as does incorporating dynamic generation of
instances as used in \Eff{}. Consequently, \Eff{} still uses the value
restriction. Our contribution is to identify a larger class of
languages in which effects and polymorphism coexist naturally.

For type-system cognoscenti, these results may not come as a complete
surprise. First, using effect systems to ensure soundness has been
proposed~\citep{leroy-weis-polymorphic-type-inference-and-assignment}
before Wright's value restriction. Second, even if we consider the
non-effect-annotated safety result, we do not believe the type system
can encode the problematic effects: local reference cells and
continuations. Nonetheless, previous solutions require a
\emph{specialised}, and sometimes subtle, type system. In the
algebraic setting, adding polymorphism to existing systems is
strikingly natural.

This result arose as part of a broader (denotational) semantic
investigation of effects and polymorphism, which does not yet account
for reference cells. We hope that an algebraic understanding of
locality~\citep{staton:instances-of-computational-effects:an-algebraic-perspective,fiore-staton:substitution-jumps-and-algebraic-effects}
and scope and polymorphic
arities~\citep{wu-schrijvers-hinze:effect-handlers-in-scope} will
explain the interaction between reference cells and polymorphism.
The robustness of type safety leads us to believe standard extensions,
such as type inference, principal types, and impredicative and row
polymorphism will not pose problems. The latter is particularly
interesting, as it can serve as an effect system with effect
variables~\citep{lindley-cheney:row-based-effect-types-for-database-integration,leijen:koka,pretnar:inferring-algebraic-effects}.

We want to investigate the expressive difference between effect
handlers and delimited control, and polymorphism forms another
comparison axis.  We defer a thorough comparison, as there are several
notions of delimited control (shift, shift0, with or without
answer-type modification) and several proposals for polymorphic type
systems
\citep{asai-kameyama:polymorphic-delimited-continuations,gunter-remy-riecke:a-generalization-of-exceptions-and-control-in-ml,kiselyov-shan-sabry:delimited-dynamic-binding},
and as delimited control is subtle. That said, there are two immediate
points of comparison between delimited control and effect handlers.

First, \citeauthor{kiselyov-shan-sabry:delimited-dynamic-binding}'s
translation of dynamic scope into delimited control requires some
complication in order to preserve the type. This complication is
caused by their effect system for delimited control tracking, the
return type of the computation enclosed by the nearest rebinding. When
an access to a dynamically scoped cell escapes the current binding in
scope the type expected in the nearest rebinding may change, resulting
in a type error of their translated program. The example on
page~\pageref{dyn-state-example} demonstrates such a shift from a
function type to an integer type. In contrast, our effect system only
tracks the local type for each effect operation, and the translation
from dynamically scoped state to effect handlers extends smoothly to
types.

Second, these systems include a form of a purity restriction or value
restriction. As a consequence, they cannot type purely functional
computations like the final example in the Evaluation
Subsection~\ref{subsec:evaluation}. In contrast, the type system
proposed here allows unrestricted Hindley-Milner polymorphism in both
purely functional and effectful code.

\paragraph{Acknowledgments}
\emph{Omitted for peer review.}

\bigskip
\bibliographystyle{jfp}
\bibliography{polymorphic-eff}

\end{document}